\RequirePackage[l2tabu, orthodox]{nag}
\documentclass[a4paper,12pt,nofootinbib]{article}
\usepackage[english]{babel} 
\usepackage[T1]{fontenc}
\usepackage{float}
\usepackage{setspace}
\usepackage{appendix}
\usepackage{parcolumns}
\usepackage{cite}
\usepackage[OT1]{fontenc}
\usepackage{type1cm}
\usepackage{url}
\usepackage{subfigure}
\usepackage{braket}
\usepackage{graphicx}
\usepackage{sidecap}
\usepackage{xspace}
\usepackage{tikz}
\usepackage{blkarray}
\usepackage{indentfirst}
\usepackage[mathscr]{eucal}

\usepackage{booktabs}
\usepackage{indentfirst}
\usepackage{bm}
\usepackage{dcolumn}
\usepackage{enumerate}
\usepackage[utf8]{inputenc}

\usepackage{amsfonts}
\usepackage{amsthm}
\usepackage{mathrsfs}
\usepackage{graphicx}
\usepackage{multirow}
\usepackage{ulem}	
\usepackage{comment}
\usepackage{amsbsy,amssymb,amsmath,amsmath,mathtools}
\usepackage{bookmark}

\usepackage{nccmath}   

\makeatletter
\@addtoreset{equation}{section}

\makeatother

\def\det{{\rm det}}

\newcommand{\vs}[1]{\vspace{#1 mm}}

\def\be{\begin{equation}}
\def\ee{\end{equation}}

\def\cL{{\mathcal L}}

\def\cP{{\mathcal P}}

\def\cO{{\mathcal O}}

\newcommand{\dd}{\mathrm{d}}

\newcommand{\trQ}[2]{\mathrm{tr}_{(#1#2)}Q}
\newcommand{\divQ}[1]{\mathrm{div}_{(#1)}Q}

\newcommand{\trdivQ}[1]{\mathrm{tr}\mathrm{div}_{(#1)}Q}
\newcommand{\Hquad}{\hspace{.25em}} 
\definecolor{monza}{HTML}{CF000F}
\definecolor{bostonuniversityred}{rgb}{0.8, 0.0, 0.0}

\newcommand{\bae}[1]{\begin{align} #1 \end{align}}
\newcommand{\multi}[1]{\begin{multline} #1 \end{multline}}
\newcommand{\beae}[1]{\begin{equation}\begin{aligned} #1 \end{aligned}\end{equation}}

\newcommand{\rpm}{m_\mathrm{P}}
\newcommand{\diag}{\mathrm{diag}}

\usepackage{physics}
\newcommand{\calL}{\mathcal{L}}
\usepackage{hyperref}
\hypersetup{backref=true,       
pagebackref=true,               
hyperindex=true,                
colorlinks=true,                
breaklinks=true,                
urlcolor= bostonuniversityred,                
linkcolor= bostonuniversityred,                
bookmarks=true,                 
bookmarksopen=false,
citecolor=bostonuniversityred,
linkcolor=bostonuniversityred,
filecolor=bostonuniversityred,
citecolor=bostonuniversityred,
linkbordercolor=bostonuniversityred
}

\begin{document}

\vs{3}
{\centering
{\large\bfseries Some simple theories of gravity with propagating nonmetricity}
\vs{8}

{\large
Yusuke Mikura${}^{\dagger *}$\footnote{e-mail address: mikura.yusuke.s8@s.mail.nagoya-u.ac.jp}
\quad
Roberto Percacci${}^*$\footnote{e-mail address: percacci@sissa.it}
}
\medskip

${}^\dagger${Department of Physics, Nagoya University,\\
Furo-cho Chikusa-ku, Nagoya 464-8602, Japan}\\
${}^*${International School for Advanced Studies, via Bonomea 265, I-34136 Trieste, Italy}\\
${}^*${INFN, Sezione di Trieste, Italy}
\bigskip
\par}
{\narrower
{\bfseries Abstract.}
We investigate symmetric Metric-Affine Theories of Gravity with a Lagrangian containing all operators of dimension up to four that are relevant to free propagation in flat space. Complementing recent work in the antisymmetric case, we derive the conditions for the existence of a single massive particle with good properties, in addition to the graviton.
}

\section{Introduction}
\label{sec:Intro}

Metric-Affine theories of Gravity (MAGs) are generalized
theories of gravity with an independent metric $g_{\mu\nu}$
and connection $A_\lambda{}^\mu{}_\nu$.
The connection transforms in a non-homogeneous way
and cannot be said to have any symmetry properties, but the difference 
\bae{
\phi_\lambda{}^\mu{}_\nu=A_\lambda{}^\mu{}_\nu-\mit\Gamma_\lambda{}^\mu{}_\nu ~,
\label{distortion}
}
is a tensor, and it makes sense to ask whether it is symmetric or antisymmetric in some indices.
Normally we decompose tensors in their symmetric and antisymmetric parts
relative to one couple of indices,
but three-index tensors have the property that they can be decomposed uniquely into two tensors where one of those is symmetric in the first and third index and the other is antisymmetric in the second and third:
\be
\phi_\lambda{}^\mu{}_\nu=L_\lambda{}^\mu{}_\nu
+K_\lambda{}^\mu{}_\nu\ ,\qquad
L_{\lambda\mu\nu}=L_{\nu\mu\lambda}\ ,\qquad
K_{\lambda\mu\nu}=-K_{\lambda\nu\mu}\ .
\ee
It therefore makes sense to classify MAGs into symmetric and antisymmetric theories,
where $K_{\lambda\mu\nu}=0$ in the first case and $L_{\lambda\mu\nu}=0$ in the second.
The tensor $L$ can be written as
\bae{
L_{\lambda\mu\nu}= \frac{1}{2}\left(Q_{\lambda\mu\nu}+Q_{\nu\mu\lambda}-Q_{\mu\lambda\nu}\right)~,
}
where $Q_{\lambda\mu\nu}$, symmetric in the last two indices, is the nonmetricity
of the connection $A$, defined by
\be
Q_{\lambda\mu\nu}=-D_\lambda g_{\mu\nu}
\equiv -\partial_\lambda g_{\mu\nu}
+A_\lambda{}^\rho{}_\mu g_{\rho\nu}
+A_\lambda{}^\rho{}_\nu g_{\mu\rho} ~.
\label{nonmet}
\ee
Similarly, the contortion tensor $K$ can be written in terms of the torsion $T$,
a tensor that is antisymmetric in the first and third index.
Thus, symmetric MAGs have nonmetricity but no torsion, 
and antisymmetric MAGs have torsion but no nonmetricity.

This classification holds at the kinematical level,
independent of any choice of Lagrangian.
Let
\bae{
F_{\rho\sigma}{}^\mu{}_\nu=
\partial_\rho A_\sigma{}^\mu{}_\nu
-\partial_\sigma A_\rho{}^\mu{}_\nu
+A_\rho{}^\mu{}_\lambda A_\sigma{}^\lambda{}_\nu
-A_\sigma{}^\mu{}_\lambda A_\rho{}^\lambda{}_\nu ~,
}
be the curvature tensor of $A$.
If we restrict ourselves to Lagrangians containing operators of dimension two,
we have only the scalar $F_{\mu\nu}{}^{\mu\nu}$
and terms quadratic in $T$ and $Q$.
Disregarding the coupling to matter,
generically the equations of motion will imply that $Q=0$ and $T=0$,
and the theory will be equivalent to General Relativity (GR).
We call these Palatini-like theories (they include Einstein-Cartan theory and others).
We are interested in the case when the tensor $\phi$
(or equivalently $L$ and $K$) contains propagating degrees of freedom.
For that, the Lagrangian must contain terms of dimension four.
A three-index tensor contains many degrees of freedom
whose propagation depends on the choice of coefficients in the Lagrangian.
Generically, some of these degrees of freedom will have pathological propagation,
being either ghosts (negative residue at the pole) or tachyons (superluminal propagation).
The number of potentially propagating degrees of freedom
and of Lagrangian coefficients determining their properties is so large that even this first step in the analysis,
that in other theories is quite straightforward, becomes a complicated task.
Some relevant early analyses in the subclass of
Poincar\'e Gauge Theories (PGTs)
\footnote{By that we mean antisymmetric MAGs with Lagrangian quadratic
in $F$ and $T$.}
were done in 
\cite{neville,Sezgin:1979zf,sezgin2,Lin:2018awc,Percacci:2020ddy}.
Some discussions in more general MAGs are contained in 
\cite{Marzo:2021esg,Marzo:2021iok,Baldazzi:2021kaf}.
In order to approach this issue in a somewhat systematic way,
in \cite{mnp} we have constructed all the cases of antisymmetric MAGs
that are free of ghosts and tachyons, propagate
only the massless graviton and another massive degree of freedom,
and have no gauge symmetries in addition to diffeomorphism.
The analysis provided several examples of ghost- and tachyon-free theories
going beyond the PGT subclass.
In this paper we will complete this task by performing the same analysis
for symmetric MAGs.

In Section~\ref{sec:lags} we write the general Lagrangian involving terms of dimension
up to four (but neglecting interactions)
and we briefly review the technicalities of the spin projector formalism.
The main results are contained in Section~\ref{sec. Healthy symmetric MAGs},
that is further subdivided into an analysis of totally symmetric
and hook symmetric MAGs.
Section~\ref{sec. Conclusions} contains a brief discussion and conclusions.

\section{Lagrangians}
\label{sec:lags}

Symmetric MAGs have been much less studied than antisymmetric ones.
Let us recall that if we think of a symmetric MAG as theory of 
an independent metric and connection,
the Lagrangian is constructed with $Q$, $F$ and their covariant
derivatives $DQ$, $DF$ etc., where $D=\partial+A$.
We call this the Cartan form of the theory.
Alternatively we can use (\ref{distortion}) to change variables
from $g$ and $A$ to $g$ and $L$, or equivalently $g$ and $Q$.
We call this the Einstein form of the theory.
The Lagrangian is then constructed with $Q$, $R$ and their covariant
derivatives $\nabla Q$, $\nabla R$ etc, 
where $\nabla=\partial+\mit\Gamma$ is the covariant derivative formed with the
Levi-Civita connection and $R$ is the curvature of the Levi-Civita connection
(i.e. the Riemann tensor).
Note that in this way of looking at it, $Q$ is not defined by (\ref{nonmet}),
but is rather an independent dynamical variable,
so in Einstein form one can think of symmetric MAG
as a metric theory of gravity coupled to a symmetric tensor $Q$
that can be thought of as a matter field.
In this paper, as in \cite{mnp}, we shall work in the Einstein form,
that is easier to analyze and understand.
Thus the Lagrangian is 
\bae{
\calL_{\mathrm{E}}=\calL_{\mathrm{E}} [g_{\mu\nu}, Q_{\alpha\beta\gamma}] ~.
}
See Refs.~\cite{Vitagliano:2010sr,Vitagliano:2013rna} for some general discussion
of the dynamics of MAG coupled to other matter.
Since here we are only interested in the propagators,
we neglect inteactions with matter as well as self-interactions.
This means that in $\calL_{\mathrm{E}}$ we do not consider any terms with more than two powers
of $Q$ and $R$.

\subsection{Dimension-two terms}

Out of $g$ and $Q$ one can construct six dimension-two operators. One of them is the Einstein--Hilbert term providing the propagator for the massless spin-2 graviton, and other five operators correspond to the mass terms of the nonmetricity, where the Lagrangian is 
\bae{\label{Eq. Einstein mass Lagrangian}
\cL^{(2)}_{\mathrm{E}}=-\frac12\left[- \rpm^2 R+\sum_{i=1}^{5} m^{QQ}_i M^{QQ}_i\right] ~,
}
with
\bae{\label{Eq. Einstein mass for Q}
M^{QQ}_1 &\, = Q^{\rho\mu\nu} Q_{\rho\mu\nu} ~,
\quad
M^{QQ}_2= Q^{\rho\mu\nu} Q_{\nu\mu\rho} ~,
\quad
M^{QQ}_3= \trQ23^{\mu} \trQ23_\mu ~,
\nonumber \\
M^{QQ}_4 &\, = \trQ12^{\mu} \trQ12_\mu ~,
\quad
M^{QQ}_5= \trQ23^{\mu} \trQ12_\mu ~.
}
Here $\rpm$ denotes the reduced Planck mass and $\tr_{(ij)}Q^a$ is the trace 
of nonmetricity on the $i$-th and $j$-th index.

\subsection{Dimension-four terms}

Leaving aside cubic and quartic interactions,
the dimension-four operators in the symmetric MAGs can be divided into three categories
\bae{
\cL_{\mathrm{E}}^{(4)}=-\frac12\left[\sum_i b^{RR}_i H^{RR}_i
+\sum_i b^{RQ}_i H^{RQ}_i
+\sum_i b^{QQ}_i H^{QQ}_i
\right] ~.
}
The first sum collects the higher curvature terms for the graviton, explicitly given by
\bae{
H^{RR}_1 & = R_{\mu\nu\rho\sigma}R^{\mu\nu\rho\sigma} ~,
\qquad
H^{RR}_2 = R_{\mu\nu}R^{\mu\nu} ~,
\qquad
H^{RR}_3 = R^2 ~,
\label{RR}
}
and the second one is a collection of couplings between the curvature and nonmetricity
\begin{equation}
    \begin{aligned}
H^{RQ}_{4} &= R^{\alpha\beta} \divQ1_{\alpha\beta} ~, &
\quad
H^{RQ}_{5} &= R^{\alpha\beta} \divQ2_{\alpha\beta} ~, &
\\
H^{RQ}_{6} &= R \, \trdivQ1 ~, &
\quad
H^{RQ}_{7} &= R \, \trdivQ2 ~. &
\label{HRQ}
\end{aligned}
\end{equation}
The last sum includes what may be called kinetic terms of nonmetricity
\begin{equation}
\begin{aligned}
H^{QQ}_{1} & = \nabla^\alpha Q^{\beta\gamma\delta} \nabla_\alpha Q_{\beta\gamma\delta} ~, & 
\Hquad
H^{QQ}_{2} & = \nabla^\alpha Q^{\beta\gamma\delta} \nabla_\alpha Q_{\gamma\beta\delta} ~, &
\\
H^{QQ}_{3} & = \nabla^\alpha\trQ12^\beta \nabla_\alpha\trQ12_\beta ~, & 
\Hquad
H^{QQ}_{4} & = \nabla^\alpha\trQ23^\beta \nabla_\alpha\trQ23_\beta  ~, &
\\
H^{QQ}_{5} & = \nabla^\alpha\trQ12^\beta \nabla_\alpha\trQ23_\beta  ~, &
\Hquad
H^{QQ}_{6} & = \divQ1^{\alpha\beta}\divQ1_{\alpha\beta} ~, & 
\\
H^{QQ}_{7} & = \divQ2^{\alpha\beta}\divQ2_{\alpha\beta} ~, &
\Hquad
H^{QQ}_{8} & = \divQ2^{\alpha\beta}\divQ2_{\beta\alpha} ~, &
\\
H^{QQ}_{9} & = \divQ1^{\alpha\beta}\divQ2_{\alpha\beta}  ~, &
\Hquad
H^{QQ}_{10} & = \divQ2^{\alpha\beta}\nabla_\alpha\trQ12_\beta ~, &
\\
H^{QQ}_{11} & = \divQ2^{\alpha\beta}\nabla_\alpha\trQ23_\beta ~, &
\Hquad
H^{QQ}_{12} & = \divQ2^{\alpha\beta}\nabla_\beta\trQ12_\alpha ~, &
\\
H^{QQ}_{13} &= \divQ2^{\alpha\beta}\nabla_\beta\trQ23_\alpha ~, &
\Hquad
H^{QQ}_{14} &= (\trdivQ1)^2 ~, &
\\
H^{QQ}_{15} &= (\trdivQ2)^2 ~, &
\Hquad
H^{QQ}_{16} &= \trdivQ1\, \trdivQ2 ~.
\label{HQQ}
\end{aligned}
\end{equation}
Here $\mathrm{div}_{i}Q_{ab}$ is the divergence of $Q$ on the $i$-th index
formed with $\nabla$, defined by 
\bae{
    \nabla = \partial +\Gamma ~.
    }

\subsection{Linearization}
\label{sec:linac}
In order to study the healthy propagation, we linearize the Einstein form of the theory around Minkowski space.~\footnote{For the metric signature, we use $\eta_{\mu\nu}=\diag\, (-1,1,1,1)$.} For the metric, we expand as
\bae{
g_{\mu\nu}=\eta_{\mu\nu}+h_{\mu\nu} ~,
}
while the nonmetricity $Q$ is treated as a fluctuation since the vacuum expectation value of the nonmetricity is zero. By Poincar\'e invariance, the linearized Lagrangian in the fourier space takes the form
\bae{
S^{(2)} = \frac12\int \frac{\dd^4q}{(2\pi)^4}
\,\Phi^T (-q)\,\cO\ \Phi (q) ~,
\label{linC}
}
where $\Phi=(Q, h)^T$ and $\cO$ is the kinetic operator written with the metric $\eta_{\mu\nu}$ and four-momentum $q^\mu$.

The fields $Q$ and $h$ can be decomposed into irreducible representations of the rotation group $O(3)$ (see table (\ref{irreps}) for the list of such irreducible representations). Those representations are labelled by spin $J$, parity $\cP$, an additional index $i$ to distinguish multiple copies of the same representation. Each representation can be singled out by means of projectors $P_{ii}(J^\cP)$, and mapping between representations with the same spin and parity but different $\{i, j\}$ indices is implemented by intertwiners $P_{ij}(J^\cP)$.~\footnote{Throughout this paper the projectors and intertwiners are collectively called projectors unless otherwise stated.}

\begin{table}[h]
{\centering
\renewcommand{\arraystretch}{1.2}
\begin{tabular}{|c|c|c|}
\hline
    & $ts$ & $hs$ \\
\hline
$TTT$   & $3^-$, $1^-_1$ & $2^-_1$, $1^-_2$ \\
\hline
$TTL+TLT+LTT$   & $2^+_1$, $0^+_1$   & - \\
\hline
$\frac32 LTT$    & - & $2^+_2$, $0^+_2$ \\
\hline
$TTL+TLT- \frac12 LTT$  & - & $1^+_1$ \\
\hline
$TLL+LTL+LLT$     & $1^-_4$   &  $1^-_5$  \\
\hline
$LLL$     & $0^+_4$   &  -  \\
\hline
\end{tabular}
\quad
\begin{tabular}{|c|c|}
\hline
 & $s$ \\
\hline
$TT$ & $2^+_4$, $0^+_5$ \\
\hline
$TL$  & $1^-_7$  \\
\hline
$LL$   &  $0^+_6$  \\
\hline
\end{tabular}
\par}
\caption{$O(3)$ spin content of projection operators for the nonmetricity (left)
and graviton (right) ($ts$ = totally symmetric, $hs$ = hook symmetric).
The symbols $L$ and $T$ in the first column refer to the longitudinal and transverse projectors defined in \eqref{Eq. def of L and T}. The subscripts distinguish different flavors of the same representation. The non-consecutive numbering follows from the conventions of Ref.~\cite{Baldazzi:2021kaf}.}
\label{irreps}
\end{table}

\smallskip

These projectors are constructed with the longitudinal and transverse projectors on vectors, defined by
\bae{\label{Eq. def of L and T}
    L_\mu{}^\nu = \frac{q_\mu q^\nu}{q^2} ~, \quad T_\mu{}^\nu = \delta_\mu^\nu -L_\mu{}^\nu ~,
}
where $q^\mu$ is a four-momentum.
These projectors were introduced for two-index tensors in Refs.~\cite{rivers,barnes,aurilia}, for antisymmetric three-index tensors in Ref.~\cite{Sezgin:1979zf}, and further generalized to arbitrary three-index tensors in Ref.~\cite{Percacci:2020ddy}.

The projectors satisfy the orthonormality
\bae{
P_{ij}(J^\cP)P_{k\ell}(J^{\prime\, \cP'})
=\delta_{JJ'}\delta_{\cP \cP'}\delta_{jk}P_{i\ell}(J^\cP)\ ,
}
and the completeness
\bae{
\sum_J\sum_\cP\sum_i P_{ii}(J^\cP)=\mathbb{I} ~.
}

Using these spin projectors, the linearized action
can be rewritten as
\bae{
S^{(2)}= \frac12\int \frac{\dd^4q}{(2\pi)^4}\sum_{J Pij}\Phi^T(-q)\cdot
a_{ij}(J^\cP)\,P_{ij}(J^\cP)\cdot \Phi(q)\ ,
\label{linactSP}
}
where we have suppressed the indices carried by the fields
and by the projectors for notational clarity.
The $a_{ij}(J^\cP)$ are called coefficient matrices,
carrying all the information about the propagation and mixing between different degrees of freedom. It should be noted that the coefficient matrices may have zero-determinant, which indicates the existence of gauge symmetries. Invariance under diffeomorphism lowers by one the rank of the coefficient matrices $a(1^-)$ and $a(0^+)$ because the transformation parameter $\xi_\mu$ can be decomposed as a three-scalar and a three-vector.
The gauge redundancy is easily fixed by suppressing these rows and columns. The remaining submatrices with non-zero determinant are written by $b_{ij}(J^\cP)$. With external sources $J = (\tau, \sigma)^T$, the equation of motion for $\Phi$ gives the following saturated propagator:
\bae{
\Pi= -\frac12\int \frac{\dd^4q}{(2\pi)^4}\sum_{J Pij}J^T (-q) \cdot
b^{-1}_{ij}(J^\cP)\,P_{ij}(J^\cP)\cdot J(q) ~.
}

Finally, let us give criteria for a healthy propagation. For each propagating degree, the inverse of the coefficient matrices $b^{-1}_{ij}$ gives a pole of propagator as 
\bae{
    b^{-1}_{ij} \propto \frac{1}{q^2 + m^2} ~,
}
where $m^2$ is written by the EFT parameters such as $m^{QQ}$ and $b^{QQ}$. Tachyon freedom thus requires that the pole mass is positive definite:
\bae{
m^2>0 ~.
}
In order not to have ghost degrees of freedom, the residue of the propagator at the pole
must also be positive:
\bae{
\eval{\mathrm{Res}\, [\Pi]}_{q^2=-m^2}>0 ~.
}
As explained in \cite{mnp}, in the evaluation of the residue one has to
take into account the number of longitudinal indices carried by the
degree of freedom in question, because each longitudinal index
is contracted with a metric that evaluates to $-1$.
(The metric evaluates to $+1$ for transverse indices. This is due to our choice
of signature $(-,+,+,+)$ and would be opposite for signature $(+,-,-,-)$,
as used for example in \cite{Sezgin:1979zf}).
The number of such indices can be read off Table~\ref{irreps}.
We shall highlight this point in the following examples.

\section{Healthy symmetric MAGs}\label{sec. Healthy symmetric MAGs}

We now look for examples of symmetric MAGs propagating only the graviton and one other 
massive degree of freedom in the nonmetricity. In order to avoid ghost degree associated with the metric tensor we set all the coefficients of the quadratic curvature $b^{RR}_i$ to be zero.~\footnote{We could allow $b^{RR}_3\not=0$ since it does not affect the propagation of the spin-two graviton.} Thus, in all cases listed below, the propagator of the graviton is the standard one from General Relativity. We also assume that the diffeomorphism is the only gauge symmetry. This means that once we remove rows and columns labelled by 7 from $a(1^-)$
and rows and columns labelled by 6 from $a(0^+)$, the remaining submatrices are nondegenerate,
and also that all the other $a$-matrices are nondegenerate. 

The nonmetricity can be decomposed into totally symmetric part and its complement dubbed hook symmetric part. These correspond to the two columns in Table \ref{irreps}.
In the next two subsections, we assume as a kinematical constraint that $Q$ has either one of these symmetries,  and study ghost- and tachyon-free conditions for single-state propagation. 

\subsection{Totally symmetric case}\label{Sec. Totally-symmetric MAG}

Here we restrict ourselves to totally symmetric nonmetricity. 
In this case several operators that are independent in a general symmetric MAG
become related:
\bae{
    M^{QQ}_2 &\, = M^{QQ}_1 ~, \quad M^{QQ}_5 = M^{QQ}_4 = M^{QQ}_3 ~, 
    \\
    H^{QQ}_2 &\, =H^{QQ}_1 ~,\quad  
    H^{QQ}_5 = H^{QQ}_4 = H^{QQ}_3 ~, \quad  
    H^{QQ}_9 = H^{QQ}_8 = H^{QQ}_7 = H^{QQ}_6 ~, 
    \nonumber\\
    H^{QQ}_{13} &\, = H^{QQ}_{12} = H^{QQ}_{11} = H^{QQ}_{10} ~,  
    \quad 
     H^{QQ}_{16} = H^{QQ}_{15} = H^{QQ}_{14} ~,
    \\
    H^{RQ}_7 &\, = H^{RQ}_6 ~, \quad H^{RQ}_5 = H^{RQ}_4 ~.
}
Making use of these relations, without loss of generality, the Lagrangian for the totally symmetric nonmetricity can be written as
\multi{\label{Eq. Totally-symmetric MAG}
\cL_{\mathrm{E}} = -\frac12  \left[-\rpm^2 R 
+ m^{QQ}_1 M^{QQ}_1 
+ m^{QQ}_3 M^{QQ}_3
+ b^{RQ}_4 H^{RQ}_4
+ b^{RQ}_6 H^{RQ}_6
\right.
\\
\left.
+ b^{QQ}_1 H^{QQ}_1
+ b^{QQ}_3 H^{QQ}_3
+ b^{QQ}_6 H^{QQ}_6
+ b^{QQ}_{10} H^{QQ}_{10}
+ b^{QQ}_{14} H^{QQ}_{14}
\right] ~.
}
As described in the previous section, all the information about propagation is imprinted in the coefficient submatrices $b_{ij}$ after removing the gauge redundancy related to the diffeomorphism.
In the totally symmetric case these nondegenerate $b$-submatrices are
\be
b(2^+)=a(2^+)_{\{1,4\},\{1,4\}}\ ,\quad
b(1^-)=a(1^-)_{\{1,4\},\{1,4\}}\ ,\quad
b(0^+)=a(0^+)_{\{1,4,5\},\{1,4,5\}}\ .
\ee 
(Here the indices refer to the number of the irreducible representation,
not to the row and column.)
They are explicitly given by
{\small
\bae{\label{Eq. b-matricies for totally-symmetric}
    b(3^-) &= -b^{QQ}_1 q^2 - m^{QQ}_1 ~,
\\
    b(2^+) &=\pmqty{-m^{QQ}_1-\left(b^{QQ}_1+\tfrac13 b^{QQ}_6\right) q^2 & \frac{i}{4\sqrt{3}} b^{RQ}_4 q^3 \\ \frac{i}{4\sqrt{3}} b^{RQ}_4 q^3  &  -\frac{\rpm^2}{4} q^2} ~,
 \\
b(1^-) &=\pmqty{-m^{QQ}_1\!\!-\tfrac53 m^{QQ}_3\!\!
-\left(b^{QQ}_1+\tfrac53 b^{QQ}_3\right) q^2
 &  -\frac{\sqrt{5}}{3}m^{QQ}_3 -\frac{\sqrt{5}}{6} \left(2b^{QQ}_3+b^{QQ}_{10}\right)q^2
 \\   -\frac{\sqrt{5}}{3}m^{QQ}_3 -\frac{\sqrt{5}}{6} \left(2b^{QQ}_3+b^{QQ}_{10}\right)q^2
 & -m^{QQ}_1\!\!+\tfrac13 m^{QQ}_3\!\!
-\frac{1}{3}\!\left(\!3b^{QQ}_1\!\!+b^{QQ}_3\!\!+2b^{QQ}_6\!\!+b^{QQ}_{10}\right)\!q^2}~,
\\
b(0^+)&=
\pmqty{ 
- m^{QQ}_1\!\! - m^{QQ}_3\!\! 
-\left(b^{QQ}_{1/3/14}+\tfrac13 b^{QQ}_6 \right)q^2
&  - m^{QQ}_3 - \left(b^{QQ}_{3/14} + \tfrac12b^{QQ}_{10}\right)q^2
& \frac{i}{4\sqrt{3}}\left(\!b^{RQ}_4\!\!+\!6 b^{RQ}_6\right)\!q^3
\\   
- m^{QQ}_3 - \left(b^{QQ}_{3/14} +\tfrac12 b^{QQ}_{10}\right)q^2
& -m^{QQ}_1\!\!-m^{QQ}_3\!\!
-b^{QQ}_{1/3/6/10/14}q^2
& \frac{\sqrt{3} i}{4}\left(\!b^{RQ}_4\!\! +\! 2 b^{RQ}_6\right)\!q^3
\\
\frac{i}{4\sqrt{3}}\left(b^{RQ}_4\!\!+6 b^{RQ}_6\right)\!q^3
& \frac{\sqrt{3} i}{4}\left(b^{RQ}_4\!\! + 2 b^{RQ}_6\right)\!q^3
&  \frac{\rpm^2}{2}q^2
}~,
}
}
where we used the shorthand $b^{QQ}_{A/B\ldots}=b^{QQ}_A+b^{QQ}_B+\ldots$.

\subsubsection{Propagation of $3^-$ state}
Let us first study the case where only the massive $3^-$ state propagates. 
To inhibit the propagation of the other modes
we have to set to zero the coefficients of $q^4$ and higher in the determinants
of $b(2^+)$ and $b(0^+)$, 
and the coefficients of $q^2$ and higher in the determinant of $b(1^-)$.
It is easy to check that this leads uniquely
to the conditions
\bae{
m^{QQ}_3 &\, =-\frac{1}{2}m^{QQ}_1 ~, \quad b^{RQ}_4=b^{RQ}_6=0 ~,
\nonumber\\
\label{Eq. parameter 3^-}
b^{QQ}_3 &\, =-\frac{2}{3}b^{QQ}_1 ~, \quad b^{QQ}_6=-3b^{QQ}_1 ~,\quad b^{QQ}_{10}=2 b^{QQ}_1 ~, \quad b^{QQ}_{14}=\frac{1}{6}b^{QQ}_1~, 
}
and the kinetic matrix for the $3^-$ state becomes
\bae{
b(3^-)= - b^{QQ}_1 \left(q^2+\frac{m^{QQ}_1}{b^{QQ}_1}\right) ~.
\label{3minusp}
}
Now we note that, with the parameter set~\eqref{Eq. parameter 3^-}, we have $\det[b(1^-)]=0$ and $\det[b(0^+)]=0$, in contradiction to the assumption that $b$ are
nondegenerate. 
This is due to the existence of an additional gauge invariance,
so that the conditions we are imposing are not enough to inhibit all propagation.
In fact, the matrices $a(1^-)$ and $a(0^+)$ have one zero eigenvalue
and one non-vanishing eigenvalue equal to
\bae{
    b(1^-)= \frac{2 b^{QQ}_1}{3} \left(q^2-\frac{3m^{QQ}_1}{2b^{QQ}_1}\right)~, \quad 
    b(0^+)= b^{QQ}_1 \left(q^2-\frac{m^{QQ}_1}{b^{QQ}_1}\right) ~,
}
respectively ($ a(0^+)$ also has the usual eigenvalue $\tfrac12 m_P^2 q^2$
related to the massless graviton).
Thus we conclude that it is impossible to have a healthy propagation of the spin $3^-$ 
and graviton alone.

Since the $3^-$ state can be extracted by a projector of type $TTT$, 
its ghost- and tachyon-free conditions are immediately evident from (\ref{3minusp}):
\bae{
    b^{QQ}_1 >0 ~, \quad m^{QQ}_1>0 ~.
}
With these conditions both the $1^-$ and the $0^+$ states
are tachyonic.
Furthermore,  since the $1^-$ state is of type $TTT/TLL$,
the condition $b^{QQ}_1 >0$ implies that it is a ghost,
whereas the $0^+$ state (that is of type $TTL/LLL$) is not. 
The propagation of massive spin 3 is interesting also for other reasons and will be the subject of a separate investigation.

\subsubsection{Propagation of $2^+$ state}
We impose the same conditions of the previous case,
except that we allow the coefficient of $q^4$ in $\det[a(2^+)]$ to be nonzero,
but put $b^{QQ}_1=0$ to stop the propagation of the $3^-$ state.
This set of conditions has the unique solution
\bae{
    m^{QQ}_3 &\, =-\frac{1}{2}m^{QQ}_1 ~, \quad b^{RQ}_4=b^{RQ}_6=0
\nonumber\\
    b^{QQ}_1 &\, = 0 ~, 
    \quad 
    b^{QQ}_{3} =\frac{1}{18}b^{QQ}_6 ~, 
    \quad
    b^{QQ}_{10}=-\frac{2}{3}b^{QQ}_6  ~, 
    \quad 
    b^{QQ}_{14} =-\frac{1}{18}b^{QQ}_6 ~.
}
The coefficient governing the propagation of the massive $2^+$ state is
\bae{
    \mathrm{b}(2^+)_{11}= -\frac{1}{3}b^{QQ}_6\left(q^2 +\frac{3m^{QQ}_1}{b^{QQ}_6}\right) ~,
}
that (using that $2^+$ state is of type $TTL$) gives the ghost- and tachyon-free conditions 
\bae{
    b^{QQ}_6<0 ~, \quad m^{QQ}_1<0 ~.
}
As in the previous case, the conditions on the parameters lead to gauge symmetries
in the $1^-$ and $0^+$ sectors.
Those matrices have zero determinant and must be further reduced.
The remaining nonzero eigenvalues of these matrices are 
\bae{
    b(1^-)=-\frac{5}{9}b^{QQ}_6\left(q^2 + \frac{9 m^{QQ}_1}{5b^{QQ}_6} \right) ~,
    \quad 
  b(0^+)=-\frac{2}{3} b^{QQ}_6 \left(q^2 + \frac{3 m^{QQ}_1}{2 b^{QQ}_6}\right) ~,
}
implying that the $1^-$ and $0^+$ states inevitably propagate on top of the $2^+$ state. 
Thus, as in the $3^-$ case, it is impossible to realize a healthy propagation of 
only the graviton and the massive $2^+$ state in totally symmetric MAGs. Incidentally we also observe that since the $1^-$ state is of type $TTT/TLL$,
the condition $b^{QQ}_6<0$ implies that it is a ghost.

\subsubsection{Propagation of $1^-$ state}
Let us now consider the propagation of the $1^-$ state. Requiring that there is no symmetry except for the diffeomorphism, the $1^-$ state can be singled out by choosing
\bae{
b^{QQ}_1= b^{QQ}_6 = b^{QQ}_{10} =b^{RQ}_4 = b^{RQ}_6 =0 ~, \quad  b^{QQ}_{14} = - b^{QQ}_3 ~,
 }
with an eigenvalue of the coefficient matrix
\bae{
    b(1^-)= - 2 b^{QQ}_3\left(q^2+\frac{m^{QQ}_1 +2 m^{QQ}_3}{2 b^{QQ}_3}\right) ~.
}
Since the $1^-$ state can come either from $TTT$ or $TLL$ projectors,
both containing an even number of $L$'s,
there is no extra minus sign and the ghost- and tachyon-free conditions are
\bae{\label{Eq. total conditions 1m}
    b^{QQ}_3 >0 ~, \quad m^{QQ}_1 +2 m^{QQ}_3>0 ~.
}

The saturated propagator for the $1^-$ state can be written in the form
\bae{
\Pi_\mathrm{{massive}} = 
\frac{1}{2}\int \frac{\dd^4q}{(2\pi)^4}\, J^{\mu} (-q)  \frac{1}{q^2+m^2} T(m^2)_{\mu\nu}  J^{\nu}(q) ~,
}
where the mass of the canonically normalized field is
\bae{
    m^2 = \frac{m^{QQ}_1 +2 m^{QQ}_3}{2 b^{QQ}_3} ~,
}
and 
\be
T(m^2)_{\mu\nu}=\eta_{\mu\nu}-\frac{q_\mu q_\nu}{m^2}
\label{Ponemass}
\ee
is the transverse projector put on shell. 
The vector source $J^\mu$ is related to the three-index source $\tau$ 
that couples linearly to $Q$ by
\bae{
J^\mu = \frac{\tau^\mu{}^\nu{}_\nu}{2\sqrt{b^{QQ}_3}} ~.
}

\subsubsection{Propagation of $0^+$ state}
As a final example of the totally symmetric MAGs, let us work on the case where only the $0^+$ state propagates. With a kinematical constraint and a requirement that there is no additional symmetry, the Lagrangian parameters are uniquely set by
\bae{
b^{QQ}_1 = b^{QQ}_3 = b^{QQ}_6 = b^{QQ}_{10} =b^{RQ}_4 = b^{RQ}_6 =0 ~.
}
Then the coefficient matrix for the $0^+$ state has an eigenvalue
\bae{
    b(0^+)=-2 b^{QQ}_{14}\left(q^2+\frac{m^{QQ}_1 + 2 m^{QQ}_3}{2b^{QQ}_{14}}\right) ~.
}
Since $0^+$ can be singled out by a projector of type either $TTL$ or $LLL$, the longitudinal projector carries extra minus sign for the ghost- and tachyon-free conditions. Thus conditions the theory should satisfy are
\bae{\label{Eq. total conditions 0p}
    b^{QQ}_{14} <0 ~, \quad m^{QQ}_1 + 2 m^{QQ}_3 <0 ~.
}

The saturated propagator for the $0^+$ is simply written as 
\bae{
\Pi_{\text{massive}}=\frac{1}{2}\int\frac{\dd^4 q}{(2\pi)^4} J(-q) \frac{1}{q^2+m^2}J(q) ~, 
}
where the mass of the canonically normalized field is 
\bae{
m^2 = \frac{m^{QQ}_1 + 2 m^{QQ}_3}{2b^{QQ}_{14}} ~.
}
In terms of the three-index source $\tau$, the scalar source $J$ is expressed as
\bae{
J=\frac{\nabla_\mu \tau^\mu{}^\nu{}_\nu}{\sqrt{2(-m^{QQ}_1-2m^{QQ}_3)}} ~.
}

\subsection{Hook symmetric case}

Hook symmetric tensors form an invariant subspace in the space of symmetric tensors,
that is complementary to the space of totally symmetric tensors.
Thus a hook symmetric nonmetricity must have vanishing totally symmetric part:
\bae{
Q^{abc}+Q^{bac}+Q^{cab}=0 ~.
}
With hook symmetry as a kinematical constraint, the number of independent operators in the Lagrangian is reduced by the following relations:
\begin{equation}
\begin{aligned}
    M^{QQ}_2 &= -\frac{1}{2}M^{QQ}_1 ~, &
    \Hquad
    M^{QQ}_4 &= \frac{1}{4}M^{QQ}_3 ~, &
    \Hquad
    M^{QQ}_5 &= -\frac{1}{2}M^{QQ}_3 ~, &
    \\
    H^{QQ}_2 &=-\frac{1}{2}H^{QQ}_1  ~, &
    \Hquad
    H^{QQ}_{4} &= 4H^{QQ}_{3}  ~, &
    \Hquad
    H^{QQ}_{5} &=-2H^{QQ}_{3}  ~,&
    \\
    H^{QQ}_{8} &= H^{QQ}_{7}+\frac{1}{2}H^{QQ}_{6}  ~, &
    \Hquad
    H^{QQ}_{9} &= -\frac{1}{2}H^{QQ}_{6} ~, &
    \Hquad
    H^{QQ}_{11} &=-2 H^{QQ}_{10}  ~, &
    \\
    H^{QQ}_{12} &=-2 H^{QQ}_{10}  ~,&
    \Hquad
    H^{QQ}_{13} &= 4 H^{QQ}_{10}  ~, &
    \Hquad
    H^{QQ}_{15} &=\frac{1}{4}H^{QQ}_{14}  ~, &
    \\
    H^{QQ}_{16} &=-\frac{1}{2}H^{QQ}_{14}  ~, &
    \Hquad
    H^{RQ}_{5} &= -\frac{1}{2}H^{RQ}_4  ~, &
    \Hquad
    H^{RQ}_{7} &= -\frac{1}{2}H^{RQ}_6  ~.
\end{aligned}
\end{equation}
Thus without loss of generality we can set all couplings to zero except
those in the following set:
\beae{
m^{QQ}_1\ ,\quad
m^{QQ}_3\ ,\quad
b^{RQ}_4\ ,\quad
b^{RQ}_6\ ,\quad
\\
b^{QQ}_1\ ,\quad
b^{QQ}_3\ ,\quad
b^{QQ}_6\ ,\quad
b^{QQ}_7\ ,\quad
b^{QQ}_{10}\ ,\quad
b^{QQ}_{14}\ .\quad
}

For generic couplings, there are no gauge symmetries beyond diffeomorphism,
and the invertible $b(J^P)$-submatrices of the coefficient matrices $a(J^P)$ are
 \bae{
&b(2^+)=a(2^+)_{\{2,4\},\{2,4\}} ~, \quad
b(2^-)=a(2^-)_{11}~, \quad 
b(1^+)=a(1^+)_{11}~, \quad 
\nonumber\\
&b(1^-)=a(1^-)_{\{2,5\},\{2,5\}}~, \quad 
b(0^+)=a(0^+)_{\{2,5\},\{2,5\}}~, \quad 
}
where the labelling follows the one of Appendix~\ref{sec:app.coefficients}.
These matrices are given by
\bae{
b(2^+)&\!=\!\pmqty{-m^{QQ}_1-\frac{1}{6}\left(6b^{QQ}_1+4b^{QQ}_6+b^{QQ}_7\right)q^2 & \frac{i}{2\sqrt{6}} b^{RQ}_4 q^3 \\ -\frac{i}{2\sqrt{6}} b^{RQ}_4 q^3 & -\frac{\rpm^2}{4} q^2} ~,
    \label{Eq. hook matrix 2p}
    \\
b(2^-)&\!=\! - m^{QQ}_1 - b^{QQ}_1 q^2  ~,
    \label{Eq. hook matrix 2m} 
    \\
b(1^+)&\!= -m^{QQ}_1 - \frac{1}{2}\left(2b^{QQ}_1 + b^{QQ}_7\right) q^2  ~,
    \label{Eq. hook matrix 1p}
   \\  
b(1^-)&\!=\!\pmqty{- m^{QQ}_1\!-\tfrac43m^{QQ}_3\!
-\!\left(\!b^{QQ}_1\!+\!\tfrac13b^{QQ}_3\!\right) q^2
 &  -\frac{4}{3\sqrt{2}}m^{QQ}_3 
-\frac{1}{3\sqrt{2}}\left(\!b^{QQ}_3+\tfrac12b^{QQ}_{10}\!\right)\!q^2
 \\  -\frac{4}{3\sqrt{2}}m^{QQ}_3 
-\frac{1}{3\sqrt{2}}\left(\!b^{QQ}_3+\tfrac12b^{QQ}_{10}\!\right)\!q^2
& - m^{QQ}_1\!-\tfrac23m^{QQ}_3\!
-\!\left(\!b^{QQ}_1\!+\tfrac16b^{QQ}_{3/10}\!
+\tfrac13 b^{QQ}_6\!+\tfrac56b^{QQ}_7\!\right)\!q^2}  ~,
\\
b(0^+)&\!=\!\pmqty{-m^{QQ}_1-2m^{QQ}_3
-\left(b^{QQ}_1+\tfrac12b^{QQ}_3 + \tfrac23b^{QQ}_6 
+\tfrac16b^{QQ}_7 +2b^{QQ}_{14}\right)q^2 
& i\frac{b^{RQ}_4+6 b^{RQ}_6}{2\sqrt{6}}q^3 
\\ -i\frac{b^{RQ}_4+6 b^{RQ}_6}{2\sqrt{6}}q^3  
& \frac{\rpm^2}{2} q^2}
    \label{Eq. hook matrix 0p}  ~.
}

We see from the coefficient matrices for the $2^+$~\eqref{Eq. hook matrix 2p} and $0^+$~\eqref{Eq. hook matrix 0p} indicate that the couplings between the nonmetricity and the graviton appear in the determinant at order $q^6$.
Thus they can only be important when more than two states propagate.
For this reason we can set $b^{RQ}_4=b^{RQ}_6=0$ 
from now on.

\subsubsection{Propagation of $2^+$ state}
Let us first consider the case of propagation of the $2^+$ state, where parameters are uniquely determined by 
\bae{
 b^{QQ}_1 &\, =b^{QQ}_7=0 ~, \quad b^{QQ}_3 =\frac{\left(b^{QQ}_{10}\right)^2}{8 b^{QQ}_6} ~, \quad b^{QQ}_{14}=-\frac{32\left(b^{QQ}_{6}\right)^2+3\left(b^{QQ}_{10}\right)^2}{96 b^{QQ}_6} ~,
    \nonumber\\
 m^{QQ}_3 &\, =-\frac{48\left(b^{QQ}_{6}\right)^2+24 b^{QQ}_{6} b^{QQ}_{10}+9\left(b^{QQ}_{10}\right)^2}{64 \left(b^{QQ}_{6}\right)^2}m^{QQ}_1 ~.
}
Then the coefficient matrix for the $2^+$ state becomes
\bae{
    b(2^+)=-\frac{2}{3}b^{QQ}_6 \left(q^2 +\frac{3 m^{QQ}_1}{2b^{QQ}_6}\right) ~.
}
With an extra minus from the projector $LTT$, the ghost- and tachyon-free conditions can be read out from the kinetic coefficient as
\bae{\label{Eq. hook conditions 2p}
    m^{QQ}_1<0 ~, \quad b^{QQ}_6<0 ~.
}

The saturated propagator of a massive spin-two particle is given by
\bae{
\Pi_\mathrm{{massive}} = \frac{1}{2}\int \frac{\dd^4q}{(2\pi)^4}\, J^{\mu\nu} (-q)  \frac{1}{q^2+m^2} P(2^+,m^2){}_{\mu\nu}{}^{\rho\sigma}  J_{\rho\sigma}(q) ~,
\label{massive2+}
}
where $J_{\mu\nu}$ is a two-index source and $P(2^+,m^2)$ is the on-shell
projector defined by
\bae{
P(2^+,m^2){}_{\mu\nu}{}^{\rho\sigma}
= T(m^2)_{(\mu}{}^{(\rho} T(m^2)_{\nu)}{}^{\sigma)}
-\frac13 T(m^2)_{\mu\nu} T(m^2)^{\rho\sigma} ~,
}
with $T\left(m^2 \right){}_{\mu\nu}$ given by (\ref{Ponemass}).
In terms of the three-index source $\tau$, the two-index source $J_{\mu\nu}$ can expressed as
\bae{
J_{\mu\nu}= \sqrt{\frac{2}{- 3 m^{QQ}_1}}\left(\nabla_\lambda \tau^\lambda{}_{\mu\nu}
-\nabla_\lambda \tau_{(\mu}{}^\lambda{}_{\nu)}\right) ~.
}

\subsubsection{Propagation of $2^-$ state}
We now move on to the case of propagation of the $2^-$ state. We can extract the state by choosing
\bae{
  &  b^{QQ}_3 =\frac{-72\left(b^{QQ}_{1}\right)^2 +12 b^{QQ}_{1} b^{QQ}_{10} -\left(b^{QQ}_{10}\right)^2}{12 b^{QQ}_1} ~,
    \quad
    b^{QQ}_6 = -b^{QQ}_1 ~,
    \quad 
    b^{QQ}_7 = -2b^{QQ}_1 ~, 
    \nonumber\\
 &   b^{QQ}_{14} = -\frac{1}{4}b^{QQ}_3 ~,
    \quad
    m^{QQ}_3= \frac{-72\left(b^{QQ}_{1}\right)^2 +16 b^{QQ}_{1} b^{QQ}_{10} -\left(b^{QQ}_{10}\right)^2}{16 \left(b^{QQ}_1\right)^2} m^{QQ}_1 ~.
}
Then the resulting coefficient matrix of the $2^-$ state becomes
\bae{
    b(2^-)=- b^{QQ}_1 \left(q^2 +\frac{m^{QQ}_1}{b^{QQ}_1}\right) ~.
}
Since the projector of the $2^-$ state is a form $TTT$, the ghost- and tachyon-free conditions are trivially
\bae{\label{Eq. hook conditions 2m}
    m^{QQ}_1 >0 ~, \quad b^{QQ}_1>0 ~.
}

The saturated propagator of the $2^-$ state is written by a three-index tensor as
\bae{
\Pi_\mathrm{{massive}} = \frac{1}{2}\int \frac{\dd^4q}{(2\pi)^4}\, J^{\mu\nu\rho} (-q)  \frac{1}{q^2+m^2} P_{\text{hs}}(2^-,m^2){}_{\mu\nu\rho}{}^{\sigma\tau\lambda}  J_{\sigma\tau\lambda}(q) ~,
\label{massive2-}
}
where $P_{\text{hs}}(2^-,m^2)$ is defined by
\multi{
    P_{\text{hs}}(2^-,m^2) = P_{11}(2^-) + \frac{q^2+m^2}{m^2}\left(P_{11}(2^+)+P_{22}(2^+)-\frac{1}{2}P_{22}(0^+)\right)
\\
    +\frac{(q^2+m^2)^2}{2 m^4} P_{55}(1^-)-\frac{q^2+m^2}{\sqrt{2}m^2}\left(P_{25}(1^-)+P_{52}(1^-) \right) ~.
}
The source $J$ is related to $\tau$ as
\bae{
    J^{\mu\nu\rho} =\frac{\tau^{\mu\nu\rho}}{\sqrt{b^{QQ}_1}} ~,
}
which confirms that a positive $b^{QQ}_1$ is required.

\subsubsection{Propagation of $1^+$ state}
For a single propagation of the $1^+$ state, the parameters have to be chosen uniquely as
\bae{
 &   m^{QQ}_3= -\frac{27\left(b^{QQ}_{7}\right)^2 + 6 b^{QQ}_{7} b^{QQ}_{10} + \left(b^{QQ}_{10}\right)^2}{36 \left(b^{QQ}_7\right)^2} m^{QQ}_1 ~, 
    \nonumber \\
&   b^{QQ}_1=0 ~,
    \quad 
    b^{QQ}_3=\frac{\left(b^{QQ}_{10}\right)^2}{18 b^{QQ}_7} ~,
\quad
    b^{QQ}_6=-\frac{1}{4}b^{QQ}_7 ~,
    \quad 
    b^{QQ}_{14}=-\frac{1}{4}b^{QQ}_3 ~.
}
Then the kinetic coefficient becomes
\bae{
    b(1^+)=-\frac{b^{QQ}_7}{2} \left(q^2 +\frac{2 m^{QQ}_1}{b^{QQ}_7}\right) ~.
}
The $1^+$ state is projected out by a projector $TTL$, so that one has to be careful about the additional sign carried by a longitudinal projector. Then the ghost- and tachyon-free conditions are given by
\bae{\label{Eq. hook conditions 1p}
    m^{QQ}_1 < 0 ~, \quad b^{QQ}_7 < 0 ~.
}
One can arrive at these conditions by checking the saturated propagator. The $1^+$ state is sourced by an antisymmetric two-index tensor and the saturated propagator takes a form
\bae{
\Pi_\mathrm{{massive}} = \frac{1}{2}\int \frac{\dd^4q}{(2\pi)^4}\, J^{\mu\nu} (-q)  \frac{1}{q^2+m^2} P(1^+,m^2){}_{\mu\nu\rho\sigma}  J^{\rho\sigma}(q) ~,
\label{massive1+}
}
where 
\bae{
    P(1^+,m^2)_{\mu\nu}{}^{\rho\sigma}= \delta^{[\rho}_{[\mu} \delta^{\sigma]}_{\nu]}+2\frac{\delta{}^{[\rho}{}_{[\mu}q_{\nu]}q^{\sigma]}}{m^2} ~.
}
Here $m^2$ denotes the mass of a canonical field, which is 
\bae{
    m^2 = \frac{2 m^{QQ}_1}{b^{QQ}_7} ~.
}
The two-index source is related to the torsion source $\tau$ by
\bae{
    J_{\mu\nu}=\frac{1}{\sqrt{-2 m^{QQ}_1}}\left(\nabla_\lambda \tau_\mu{}^\lambda{}_\nu-\nabla_\lambda \tau_\nu{}^\lambda{}_\mu\right) ~,
}
showing that the ghost- and tachyon-free conditions~\eqref{Eq. hook conditions 1p} should be satisfied.

\subsubsection{Propagation of $1^-$ state}
Let us move on to the case where only the $1^-$ state propagates. Requiring all other degrees are not dynamical, the parameters are uniquely fixed by 
\bae{
    b^{QQ}_1= b^{QQ}_6= b^{QQ}_7 =b^{QQ}_{10}=0 ~, \quad b^{QQ}_{14}=-\frac{1}{4}b^{QQ}_3 ~.
}
The coefficient matrix of the state now has an eigenvalue 
\bae{
    b(1^-)= -\frac{b^{QQ}_3}{2} \left(q^2 + \frac{2(m^{QQ}_1 + 2 m^{QQ}_3)}{b^{QQ}_3}\right) ~.
}
Since a projector of the $1^-$ state is either $TTT$ or $TLL$, for a healthy propagation, the remaining parameters should satisfy
\bae{\label{Eq. hook conditions 1m}
    b^{QQ}_3 >0 ~, \quad m^{QQ}_1 + 2m^{QQ}_3 >0 ~.
}

This can be also checked by an explicit form of the saturated propagator. The $1^-$ state is sourced by a vector field $J_\mu$ as 
\bae{
\Pi_\mathrm{{massive}} = 
\frac{1}{2}\int \frac{\dd^4q}{(2\pi)^4}\, J^{\mu} (-q)  \frac{1}{q^2+m^2} T(m^2)_{\mu\nu}  J^{\nu}(q)\ ,
\label{massive1-}
}
where $T(m^2)$ is the transverse projector put on-shell and $m^2$ is the mass for a canonically-normalized field given by 
\bae{
    m^2 = \frac{2 (m^{QQ}_1 + 2 m^{QQ}_3)}{b^{QQ}_3} ~.
}
In terms of the three-index source $\tau$, the vector source can be written as
\bae{
    J_\mu = \frac{2}{3\sqrt{b^{QQ}_3}}\left(\tau_\mu{}^\nu{}_\nu -\tau_\nu{}^\nu{}_\mu\right) ~,
}
indicating that $b^{QQ}_3$ should be positive-definite.

\subsubsection{Propagation of $0^+$ state}

Let us finally work on the case where the $0^+$ state propagates. Once we require that there is no symmetry on top of the diffeomorphism, we obtain a unique set of Lagrangian parameters
\bae{
    b^{QQ}_1= b^{QQ}_3=b^{QQ}_6=b^{QQ}_7=b^{QQ}_{10}=0 ~.
}
The propagation of the massive $0^+$ state is governed by
\bae{
    b(0^+)=
    -2b^{QQ}_{14}\left(q^2 +\frac{m^{QQ}_1+2m^{QQ}_3}{2b^{QQ}_{14}}\right) ~.
}
Recalling that the projector of this state takes the form $LTT$, the ghost- and tachyon-free conditions are 
\bae{\label{Eq. hook conditions 0p}
    b^{QQ}_{14} <0 ~, \quad m^{QQ}_1+2m^{QQ}_3<0 ~.
}
The saturated propagator of a massive scalar is simply given by
\bae{
\Pi_\mathrm{{massive}} = \frac{1}{2}\int \frac{\dd^4q}{(2\pi)^4}\ J(-q) \frac{1}{q^2+m^2} J(q) ~.
}
This scalar source $J$ is related to the three-index tensor as
\bae{
   J= \frac{2}{3\sqrt{-2(m^{QQ}_1+2m^{QQ}_3)}}\left(\nabla_\mu\tau^\mu{}^\nu{}_\nu -\nabla_\mu\tau_\nu{}^\nu{}^\mu\right) ~,
}
which is consistent with the obtained conditions.

\section{Conclusions}\label{sec. Conclusions}

Metric-Affine Gravity is a wide class of gravitational theories where the metric and 
a three-index tensor, which is related to the connection, are treated as independent variables.
In order to proceed systematically, we have focused on “simple” MAGs, 
meaning theories with the following properties:
\begin{itemize}
\item[1)] only one massive degree of freedom propagates in addition to the graviton,
\item[2)] there are no symmetries in addition to diffeomorphism (whether accidental or not),
\item[3)] there are no higher derivative terms for the metric field.
\end{itemize}
The task is to find all ghost- and tachyon-free theories of this type.
In Ref.~\cite{mnp} we considered this question for antisymmetric MAGs,
i.e. theories with torsion only, and in the present paper we completed the task
by considering the question in the case of symmetric MAGs,
i.e. theories with nonmetricity only.
In both cases we have catalogued all the theories with the desired properties,
thereby finding several new examples of ghost- and tachyon-free MAGs.

The procedure we followed had been advocated in \cite{Baldazzi:2021kaf}
as ``DIY MAGs'' and is constructive in nature:
one first solves the conditions on the Lagrangian parameters to inhibit the propagation of the unwanted degrees of freedom. When solutions exist, they relate certain parameters
to other, reducing the dimension of theory space.
Then one imposes that the desired degrees of freedom do not behave
pathologically, which leads in general to inequalities that constrain theory space
without reducing its dimension further.

There are several possible directions to extend this work.
We have seen in \cite{mnp} that the solutions we found in the presence of
kinematical constraints (e.g. total- or hook-antisymmetry) also exist
in the larger theory space of antisymmetric MAGs.
We expect this to be true also in the symmetric case, and in general MAGs. 

Another obvious generalization is to allow several particles to propagate simultaneously.
This may lead to useful insights, because we have seen that
demanding good propagation for one particle may lead to pathologies for another.
While a full classification of all non-pathological models is a daunting task, it is particularly interesting to investigate the possibilities where a $3^-$ or $2^+$ state propagates without any pathologies in a general setup, as we have seen in Section~\ref{Sec. Totally-symmetric MAG} that these states can not propagate on their own, at least when the kinematical constraint
of total/hook-symmetry are imposed.

Another generalization is to relax condition (3) above.
This condition is nearly implied by demanding absence of ghosts,
because this rules out Riemann squared or Ricci squared terms,
but a purely $R^2$ term would still be allowed.
We have not considered this case because it has been widely studied in the literature.
However, a more interesting case would be to look for MAGs that
are ghost- and tachyon-free {\it in spite of} the presence of
general curvature squared terms.
That such cases exist has been shown in \cite{muko} for antisymmetric MAG
and it would be interesting to examine this issue for symmetric MAG.

We have studied MAGs in what we called the Einstein form, where the
independent variables are the metric and the distortion tensor.
Much of the appeal of MAGs comes from their similarity to Yang-Mills theories,
which is only apparent in what we call the Cartan form,
where the independent variables are the metric and the connection.
These formulations are related by a change of variables and are
therefore physically equivalent, at least at classical level.
One may conjecture that perhaps in the quantum theory
the connection is a more fundamental
degree of freedom of the metric (the latter being an order parameter
signalling the Higgsing of the local $GL(4)$ symmetry \cite{percacci2,Percacci:2023rbo})
in which case the Cartan form of the theory would be
more fundamental.
This requires the exploration of quantum properties of MAG,
which is in a very preliminary stage \cite{Melichev:2023lwj}.

\bigskip

\section*{Acknowledgements}
We acknowledge the free software packages
{\tt xAct}, {\tt xTensor}, {\tt xPert}, {\tt xTras}. YM is supported by JSPS KAKENHI Grants No.~22J22254 and JSPS Overseas Challenge Program for Young Researchers.

 \break


\begin{appendix}
\section{Matrix Coefficients}
\label{sec:app.coefficients}

Recall from Table~\ref{irreps} that symmetric MAG contains the following degrees of freedom:
$3^-$, $2^+_{1,2,4}$, $2^-_1$, $1^+_1$, $1^-_{1,2,4,5,7}$ and $0^+_{1,2,4,5,6}$.
The kinetic coefficients describing the propagation and mixing of the
corresponding particles are listed below, in terms of the couplings of the
theory in the Einstein form.
The vanishing of the coefficients $a(1^-)_{k7}$, $a(0^+)_{k6}$ is due
to diffeomorphism invariance of the action.
\small 
\begin{fleqn}[0pt]
    \begin{equation}
        \begin{aligned}
    a(3^{-}) = \, -(b^{QQ}_{1} + b^{QQ}_{2})\, q^2 - ( m^{QQ}_{1} + m^{QQ}_{2})
\end{aligned}
\end{equation}
\begin{equation}
    \begin{aligned}
    a(2^{+})_{11} =& \,  -\frac{1}{3}(3 b^{QQ}_{1} + 3 b^{QQ}_{2} + b^{QQ}_{6}+b^{QQ}_{7}+b^{QQ}_{8}+b^{QQ}_{9})\, q^2 - (m^{QQ}_{1}+ m^{QQ}_{2})
\end{aligned}
\end{equation}
\begin{equation}
\begin{aligned}
    a(2^{+})_{12} =& \, -\frac{1}{6\sqrt{2}}(4 b^{QQ}_{6} -2 b^{QQ}_{7} -2 b^{QQ}_{8} + b^{QQ}_{9})\, q^2
\end{aligned}
\end{equation}
\begin{equation}
\begin{aligned}
    a(2^{+})_{14} =& \, i \frac{1}{4\sqrt{3}}(b^{RQ}_{4}+b^{RQ}_{5})\, q^3
\end{aligned}
\end{equation}
\begin{equation}
\begin{aligned}
    a(2^{+})_{22} =& \, -\frac{1}{6}(6 b^{QQ}_{1}-3 b^{QQ}_{2} +4 b^{QQ}_{6} +b^{QQ}_{7}+b^{QQ}_{8}-2b^{QQ}_{9})\, q^2 -\frac{1}{2}(2 m^{QQ}_{1} - m^{QQ}_{2})
\end{aligned}
\end{equation}
\begin{equation}
\begin{aligned}
    a(2^{+})_{24} =& \, i \frac{1}{4\sqrt{6}}(2 b^{RQ}_{4} - b^{RQ}_{5}) \, q^3 
\end{aligned}
\end{equation}
\begin{equation}
\begin{aligned}
    a(2^{+})_{44} =& \, -\frac{1}{4}(4 b^{RR}_{1} + b^{RR}_{2}) \, q^4 -\frac{\rpm^2}{4}\, q^2
\end{aligned}
\end{equation}
\begin{equation}
    \begin{aligned}
    a(2^{-})_{11} =& \, -\frac{1}{2}(2 b^{QQ}_1-b^{QQ}_2)\, q^2 -\frac{1}{2}(2 m^{QQ}_1 - m^{QQ}_2)
\end{aligned}
\end{equation}
\begin{equation}
    \begin{aligned}
    a(1^{+})_{11} =& \, -\frac{1}{2}(2 b^{QQ}_1-b^{QQ}_2+b^{QQ}_7-b^{QQ}_8)\, q^2 -\frac{1}{2}(2 m^{QQ}_1 - m^{QQ}_2)
\end{aligned}
\end{equation}
\begin{equation}
    \begin{aligned}
    a(1^{-})_{11} =& \, -\frac{1}{3}(3 b^{QQ}_1 +3 b^{QQ}_2+5b^{QQ}_3 + 5 b^{QQ}_4+5b^{QQ}_5)\, q^2 
    \\
    & \qquad -\frac{1}{3}(3 m^{QQ}_1 +3 m^{QQ}_2 + 5 m^{QQ}_3 + 5 m^{QQ}_4 + 5 m^{QQ}_5)
\end{aligned}
\end{equation}
\begin{equation}
    \begin{aligned}
    a(1^{-})_{12} =& \, -\frac{\sqrt{5}}{6}(2 b^{QQ}_3 - 4 b^{QQ}_4 - b^{QQ}_5)\, q^2 -\frac{\sqrt{5}}{6}(-4 m^{QQ}_3 + 2 m^{QQ}_4 - m^{QQ}_5)
\end{aligned}
\end{equation}
\begin{equation}
    \begin{aligned}
    a(1^{-})_{14} =& \, -\frac{\sqrt{5}}{6}(2b^{QQ}_3 + 2 b^{QQ}_4 + 2 b^{QQ}_5 + b^{QQ}_{10}+b^{QQ}_{11}+b^{QQ}_{12}+b^{QQ}_{13})\, q^2
    \\
    & \qquad
    -\frac{\sqrt{5}}{3}(m^{QQ}_3 + m^{QQ}_4 + m^{QQ}_5)
\end{aligned}
\end{equation}
\begin{equation}
    \begin{aligned}
    a(1^{-})_{15} =& \, -\frac{\sqrt{5}}{6\sqrt{2}} (2b^{QQ}_3-4 b^{QQ}_4- b^{QQ}_5+b^{QQ}_{10}+b^{QQ}_{11}-2b^{QQ}_{12}-2b^{QQ}_{13})\, q^2
    \\
    & \qquad
    -\frac{\sqrt{5}}{6 \sqrt{2}}(-4 m^{QQ}_3 +2 m^{QQ}_4 - m^{QQ}_5)
\end{aligned}
\end{equation}
\begin{equation}
    \begin{aligned}
    a(1^{-})_{17} =& \, 0 
\end{aligned}
\end{equation}
\begin{equation}
    \begin{aligned}
    a(1^{-})_{22} =& \, -\frac{1}{6}(6b^{QQ}_1-3 b^{QQ}_2 +2b^{QQ}_3 + 8 b^{QQ}_4 - 4 b^{QQ}_5) \, q^2 
    \\
    & \qquad
    -\frac{1}{6}(6 m^{QQ}_1 -3 m^{QQ}_2 +8 m^{QQ}_3 + 2 m^{QQ}_4 - 4 m^{QQ}_5)
\end{aligned}
\end{equation}
\begin{equation}
    \begin{aligned}
    a(1^{-})_{24} =& \, -\frac{1}{6}(2b^{QQ}_3 -4 b^{QQ}_4 - b^{QQ}_5 + b^{QQ}_{10} -2 b^{QQ}_{11}+ b^{QQ}_{12} -2 b^{QQ}_{13}) \, q^2 
    \\
    & \qquad
    -\frac{1}{6}(-4 m^{QQ}_3 +2  m^{QQ}_4 -m^{QQ}_5)
\end{aligned}
\end{equation}
\begin{equation}
    \begin{aligned}
    a(1^{-})_{25} =& \, -\frac{1}{6\sqrt{2}}(2b^{QQ}_3 +8 b^{QQ}_4 -4 b^{QQ}_5 + b^{QQ}_{10} -2 b^{QQ}_{11}-2 b^{QQ}_{12} +4 b^{QQ}_{13}) \, q^2 
    \\
    & \qquad
    -\frac{1}{3\sqrt{2}}(4 m^{QQ}_3 + m^{QQ}_4 - 2 m^{QQ}_5)
\end{aligned}
\end{equation}
\begin{equation}
    \begin{aligned}
        a(1^{-})_{44}  =& \, -\frac{1}{3}(3 b^{QQ}_1 + 3 b^{QQ}_2 + b^{QQ}_3 + b^{QQ}_4 + b^{QQ}_5 + 2 b^{QQ}_6 + 2 b^{QQ}_7
        \\
        & \qquad + 2 b^{QQ}_8 + 2 b^{QQ}_9 + b^{QQ}_{10} + b^{QQ}_{11} + b^{QQ}_{12} + b^{QQ}_{13}) \, q^2 
        \\
        & \qquad
        -\frac{1}{3} (3 m^{QQ}_1 +3 m^{QQ}_2 +m^{QQ}_3 + m^{QQ}_4 + m^{QQ}_5)
\end{aligned}
\end{equation}
\begin{equation}
    \begin{aligned}
    a(1^{-})_{45} =& \, -\frac{1}{6\sqrt{2}}(2 b^{QQ}_3 - 4 b^{QQ}_4 - b^{QQ}_5 + 4 b^{QQ}_6 - 2 b^{QQ}_7
    \\
    & \qquad
     - 2 b^{QQ}_8 + b^{QQ}_9 + 2 b^{QQ}_{10} - b^{QQ}_{11} - b^{QQ}_{12} - 4 b^{QQ}_{13}) \, q^2 
    \\
    & \qquad
    -\frac{1}{6\sqrt{2}}(-4 m^{QQ}_3 +2 m^{QQ}_4 - m^{QQ}_5)
\end{aligned}
\end{equation}
\begin{equation}
    \begin{aligned}
    a(1^{-})_{55} =& \, -\frac{1}{6}(6 b^{QQ}_1 - 3 b^{QQ}_2 + b^{QQ}_3 + 4 b^{QQ}_4 - 2 b^{QQ}_5 + 2 b^{QQ}_6 + 5 b^{QQ}_7
    \\
    & \qquad
    - 4 b^{QQ}_8 - b^{QQ}_9 + b^{QQ}_{10} - 2 b^{QQ}_{11} - 2 b^{QQ}_{12} + 4 b^{QQ}_{13}) \, q^2 
        \\
    & \qquad
    -\frac{1}{6}(6 m^{QQ}_1 - 3 m^{QQ}_2 + 4 m^{QQ}_3 + m^{QQ}_4 - 2 m^{QQ}_5)
\end{aligned}
\end{equation}
\begin{equation}
\begin{aligned}
    a(0^{+})_{11} =& \, -\frac{1}{3}(3 b^{QQ}_1 + 3 b^{QQ}_2 + 3 b^{QQ}_3 + 3 b^{QQ}_4 + 3 b^{QQ}_5 + b^{QQ}_6 + b^{QQ}_7 
    \\
    & \qquad
    + b^{QQ}_8 + b^{QQ}_9 + 3 b^{QQ}_{14} + 3 b^{QQ}_{15} + 3 b^{QQ}_{16}) \, q^2 
    \\
    & \qquad
    -\frac{1}{3}(3 m^{QQ}_1 + 3 m^{QQ}_2 + 3 m^{QQ}_3 + 3 m^{QQ}_4 + 3 m^{QQ}_5)
\end{aligned}
\end{equation}
\begin{equation}
\begin{aligned}
        a(0^{+})_{12} =& \, -\frac{1}{6\sqrt{2}}(-6 b^{QQ}_3 + 12 b^{QQ}_4 + 3 b^{QQ}_5 + 4 b^{QQ}_6 - 2 b^{QQ}_7 
        \\
        & \qquad
        - 2 b^{QQ}_8 + b^{QQ}_9 + 12 b^{QQ}_{14} - 6 b^{QQ}_{15} + 3 b^{QQ}_{16}) \, q^2 
        \\
        & \qquad
        -\frac{1}{6\sqrt{2}}(12 m^{QQ}_3 - 6 m^{QQ}_4 + 3 m^{QQ}_5)
\end{aligned}
\end{equation}
\begin{equation}
\begin{aligned}
            a(0^{+})_{14} =& \, -\frac{1}{2}(2 b^{QQ}_3 + 2 b^{QQ}_4 + 2 b^{QQ}_5 + b^{QQ}_{10} + b^{QQ}_{11} + b^{QQ}_{12} + b^{QQ}_{13} + 2 b^{QQ}_{14} + 2 b^{QQ}_{15} + 2 b^{QQ}_{16}) \, q^2 
            \\
            & \qquad
            -(m^{QQ}_3 + m^{QQ}_4 + m^{QQ}_5)
\end{aligned}
\end{equation}
\begin{equation}
    \begin{aligned}
                a(0^{+})_{15} =& \, i\frac{1}{4\sqrt{3}}(b^{RQ}_4 + b^{RQ}_5 + 6 b^{RQ}_6 + 6 b^{RQ}_7) \, q^3
    \end{aligned}
    \end{equation}
\begin{equation}
\begin{aligned}
            a(0^{+})_{22} =& \, -\frac{1}{6}(6 b^{QQ}_1 - 3 b^{QQ}_2 + 3 b^{QQ}_3 + 12 b^{QQ}_4 - 6 b^{QQ}_5 + 4 b^{QQ}_6 + b^{QQ}_7
            \\
            & \qquad
            + b^{QQ}_8 - 
            2 b^{QQ}_9 + 12 b^{QQ}_{14} + 3 b^{QQ}_{15} - 6 b^{QQ}_{16}) \, q^2 
            \\
            & \qquad
            -\frac{1}{6}(6 m^{QQ}_1 - 3 m^{QQ}_2 + 12 m^{QQ}_3 + 3 m^{QQ}_4 - 6 m^{QQ}_5)
\end{aligned}
\end{equation}
\begin{equation}
\begin{aligned}
            a(0^{+})_{24} =& \, -\frac{1}{2\sqrt{2}}(-2 b^{QQ}_3 + 4 b^{QQ}_4 + b^{QQ}_5 - b^{QQ}_{10}+ 2 b^{QQ}_{11} 
            \\
            & \qquad
            - b^{QQ}_{12}
            + 2 b^{QQ}_{13} + 4 b^{QQ}_{14} - 2 b^{QQ}_{15} + b^{QQ}_{16}) \, q^2 
            \\
            & \qquad
            -\frac{1}{2\sqrt{2}}(4 m^{QQ}_3 - 2 m^{QQ}_4 + m^{QQ}_5)
\end{aligned}
\end{equation}
\begin{equation}
\begin{aligned}
            a(0^{+})_{25} =& \, i\frac{1}{4\sqrt{6}}(2 b^{RQ}_4 - b^{RQ}_5 + 12 b^{RQ}_6 - 6 b^{RQ}_7) \, q^3
\end{aligned}
\end{equation}
\begin{equation}
\begin{aligned}
            a(0^{+})_{44} =& \, -(b^{QQ}_1 + b^{QQ}_2 + b^{QQ}_3 + b^{QQ}_4 + b^{QQ}_5 + b^{QQ}_6 + b^{QQ}_7 + b^{QQ}_8 
            \\
            & \qquad
            + b^{QQ}_9 + b^{QQ}_{10}+ b^{QQ}_{11} + b^{QQ}_{12} + b^{QQ}_{13} + b^{QQ}_{14} + b^{QQ}_{15} + b^{QQ}_{16}) \, q^2
            \\
            & \qquad
            - (m^{QQ}_1 + m^{QQ}_2 + m^{QQ}_3 + m^{QQ}_4 + m^{QQ}_5)
\end{aligned}
\end{equation}
\begin{equation}
\begin{aligned}
            a(0^{+})_{45} =& \, i\frac{\sqrt{3}}{4}(b^{RQ}_4 + b^{RQ}_5 + 2 b^{RQ}_6 + 2 b^{RQ}_7) \, q^3
\end{aligned}
\end{equation}
\begin{equation}
\begin{aligned}
            a(0^{+})_{55} =& \, (b^{RR}_{1} + b^{RR}_{2} + 3 b^{RR}_{3})\, q^4 + \frac{\rpm^2}{2}\, q^2
\end{aligned}
\end{equation}
\end{fleqn}

\end{appendix}

\normalsize


\end{document}